# Decarbonising the EU Power Sector: a Technological and Socio-economic Analysis and the Role of Nuclear

Maria Papadopoulou,* Roberto Passalacqua, Domenico Rossetti di Valdalbero, Elena Righi Steele

* Corresponding author: European Commission, 1049 Brussels, Belgium - Tel. +32 229-62228, Maria.PAPADOPOULOU11@ec.europa.eu.

*The views expressed are purely those of the authors and may not under any circumstances be regarded as stating an official position of the European Commission.*

**Abstract**

Low-carbon electricity is a key enabler in combating climate change. Decarbonising the power sector is now at the centre of global and European policies. As the IPCC highlights, pathways where the power sector rapidly decarbonises by 2030 have higher chances of keeping global warming below 1.5°C. The electricity sector should be fully decarbonised by 2050 to meet either the 1.5°C or 2°C targets. This means that EU policy efforts should focus on supporting a maximum reduction of emissions per unit of electricity by 2030 and net-zero emissions by 2050. Reaching these targets is one of the most pressing questions EU policymakers face today. In light of the COVID-19 crisis, EU policies should guide a cost-effective, reliable and environmentally sound transition of the power sector, benefiting EU research and innovation and its citizens. This meta-analysis provides a novel view on historical data and compares data from modelling scenarios identified in the literature. It assesses the current and future role of nuclear energy in decarbonizing the EU power sector, while reviewing socio-economic implications that could arise if limited public support nearly excludes nuclear fission electricity from the future EU power mix. This work highlights relevant socio-economic policy implications and actionable policy recommendations.

## 1 Introduction

The Paris Agreement[1] represents the global effort to combat climate change with a central aim to keep global warming this century well below 2°C above pre-industrial levels, while pursuing efforts to limit warming to 1.5°C. A recent IPCC report has however revealed that the current post-2020 pledges of the signatories deviate from the 2°C-consistent pathways and are broadly aligned with a warming of about 3°C by 2100 (Intergovernmental Panel on Climate Change, 2018) and a potential global emissions peak at 51 GtCO2e/year as early as 2025 (Kitous et al., 2017). More importantly, the report states that if the 2030 global emissions estimated under the current pathways are reached, limiting warming to 1.5°C would not be possible thereafter, even with additional steeper emissions reductions. Global emissions should start declining well before 2030 in order to avoid overshoot. Delayed action risks cost escalation, lock-in carbon-emitting infrastructure, stranded assets, and reduced flexibility in the long-term response options (Capros et al., 2014; European Commission, 2018). This message is especially important for the European Union (EU), where future dependence on large-scale deployment of carbon capture and storage (CCS) technologies is becoming increasingly uncertain: their cost-effectiveness is yet to be proven as none of the CCS projects that received EU funding has so far achieved the intended progress in demonstrating the technology at commercial scale (European Court of Auditors, 2018).

---

[1] Adopted December 12, 2015 and entered into force November 4, 2016:
https://unfccc.int/files/essential_background/convention/application/pdf/english_paris_agreement.pdf.

The IPCC report further highlights that pathways with higher chances of keeping warming this century to below 1.5°C show a rapid decarbonisation of the power sector by 2030 and around mid-century the sector should be essentially fully decarbonised to meet either the 1.5°C or the 2°C target (Intergovernmental Panel on Climate Change, 2018). Consequently, this means that the EU should target a net-zero emissions electricity by 2050 and at the same time prioritise efforts to maximize the sector's emissions reduction by 2030. Reaching these targets has become one of the most-pressing questions that EU policymakers face today and in-light of the COVID-19 crisis, it is imminent that policies guide a cost-effective, reliable, energy-efficient and environmentally-sound transition of the power sector, bringing an added value to research and innovation and society at large.

To date, the EU has set some of the most ambitious energy and climate policy frameworks (European Commission, 2016a) and aims to commit to climate-neutrality by 2050 (European Commission, 2020), an objective that is at the heart of the EU Green Deal (European Commission, 2019a). The European Commission (EC) has therefore developed a set of medium- and long-term scenarios to allow policy makers to assess the potential medium- and long-term impacts of the current policy framework and assess policy proposals for achieving the EU's climate and energy targets for 2030 and 2050 (Capros et al., 2016; E3MLab, 2019; E3MLab and IIASA, 2016). Aside from these technical reports, there exists no in-depth analysis on the economic, energy and environmental impacts for the EC medium-term scenarios, especially one that compares medium-term scenario projections to the EU's decarbonisation and energy transformation trends to date.

On the contrary, there exist comprehensive analyses for the long-term projections under the current EU policy framework as well as the projected impacts of the various long-term scenarios reflecting specific policy pathways (Capros et al., 2016, 2014; European Commission, 2018; Knopf et al., 2015). These reports include in-depth analyses of the energy and emission-related impacts by sector and by policy scenario, including assessments of some socio-economic aspects such as investment needs and energy-system costs and prices. However, the assumptions and boundary conditions of both the medium- and long-term EC scenarios reflect technology-specific energy and climate targets and consider particular policy incentives for renewable energy sources (RES), such as feed-in tariffs or green certificates, which can influence the interplay of technologies in the modelling and in turn the cost-effectiveness of the resulting power-mix. It has been shown compared to these approaches that, policy routes designed to achieve cost-effective decarbonisation of the EU power sector while supporting less technology-specific targets and excluding RES policy incentives, would yield substantially different EU power-generation mixes both in the medium- and long-term (Simoes et al., 2017). In particular, among the policy routes investigated, the one that could yield the highest impact on the cost-effective decarbonisation of the EU power-mix, was when social acceptance for nuclear electricity was assumed to be higher: lifting restrictions to the deployment of new nuclear power plants (NPP) led to a cost-effective power mix with the largest amount of electricity production by 2050 due to the significant increases in future nuclear power generation. Interestingly, the study also revealed that the policy route that could have the second highest impact on the decarbonisation of the EU power-mix would assume a higher potential for solar and wind power generation, which would require increasing the number of available RES sites and improving the reliability of transmission and distribution, factors which both reflect public acceptance to perceived technology risks, land use change and energy security concerns. These results imply that there exist socio-economic aspects and policy incentives with an exogenous influence on the EU power-mix scenarios that could have implications for the role of nuclear power and the power-system's cost-effective decarbonisation, which require further investigation.

In addition, socio-economic externalities (European Commission, 2003), such as those related to human health-related impacts (cf. toxicity or accidents), climate change, the availability of land, public acceptance, or competitiveness of EU-located electricity production compared to imports that could arise from the energy transition, are not fully captured by modelling scenarios.[2] Many studies have highlighted the wide and contrasting estimates of electricity generation costs for the various technologies and main factors identified are the inconsistent usage of terms and the differences in the cost elements assessed in the overall power system costs (Bustreo et al., 2017; D'haeseleer, 2013; NEA, 2018). These inconsistencies restrict coherent comparisons among technologies and could result in a significant over- or under-estimation of the efficiency of the power system in terms of cost, environmental and social performance, as well as its flexibility, reliability and resilience.

This work aims to complement the understanding of the policy implications of the EU energy and climate policy framework by highlighting possible emerging economic, social and environmental aspects, focusing on the growing importance of the EU power sector mid- and long-term decarbonisation. Through a meta-analysis that provides a novel view on available historical data, as well as on comparing data from modelling scenarios identified in literature, it sheds light to the central role of electricity in decarbonising the EU economy and to the current and future contribution of nuclear power as a source of low-carbon and energy-dense technology, while identifying relevant socio-economic policy implications and supporting actionable policy recommendations.

After briefly presenting the data sources and modelling scenarios (Section 2), the paper first analyses historical trends observed in the EU to date focusing on the dimensions of energy security, emissions and decarbonisation costs (Section 3). It then investigates possible medium- and long-term implications that can be identified along these dimensions from the various EC policy pathways and draws links to alternative, less-technology specific, policy pathways, where nuclear power could have a less, or more prominent role in the provision of electricity (Section 4). Some potential socio-economic implications are further presented in this section, and the following section (Section 5) provides a complementary discussion on potential policy-related implications with respect to assessing full-system costs of electricity generation technologies including those on climate change, human toxicity, and agricultural land use, depletion of energy resources, nuclear accidents and other. Finally, the main conclusions from this analysis are drawn, summarising key implications and discussing relevant policy recommendations (Section 6).

---

[2] It should be noted that macro-economic modelling was also undertaken to estimate the impacts of the scenarios on GDP growth and jobs (E3MLab, 2016; Lewney et al., 2017).

## 2 Modelling Scenarios and Data

The analysis is divided into two sections. The first (Section 3) focuses on investigating the EU decarbonisation and energy transformation trends to date based on historical data retrieved from the official, open, and databases of Eurostat, European Commission, United Nations, World Bank, International Energy Agency and Potsdam Institute for Climate Impact Research (European Commission, Directorate-General for Energy, 2020; Eurostat, 2020; Gütschow, et al., 2019; International Energy Agency, Organisation for Economic Co-operation and Development, 2020; United Nations, Department of Economic and Social Affairs, Population Division, 2019; World Bank, 2020). The data retrieved were along the dimensions energy security, emissions and decarbonisation costs and included values on energy consumption, GDP, emissions, net-imports and relevant shares across energy fuels and sectors. Since data were available from different data sources, in order to retain data consistency and be in-line with the policy frameworks and modelling results, a timeline between 1990 and 2017 using a one-year step, was selected for the total of EU-28 member states at the time the analysis was performed. Complementary data on electricity prices and costs are derived from official EC published documents (European Commission, 2019b, 2016b) and the relevant supporting study on behalf of EC (Rademaekers et al., 2018). The second section explores the EU28 decarbonisation projections from different medium- (by 2030) and long-term (by 2050) scenarios along the dimensions of energy security, emissions and decarbonisation costs in the EU28, with a particular focus on the EU power sector and the projected role of nuclear power.

First, published data from EC modelling studies are analysed (E3MLab, 2019; E3MLab and IIASA, 2016; European Commission, 2019c), which were developed as a set of medium- and long-term scenarios, in an effort to assess the potential impact of achieving the EU's climate and energy targets for 2030 and 2050. The EC scenarios represent trend projections[3], with the EU Reference Scenario 2016 (REF2016) acting as a benchmark of policy and market trends against which several policy proposals were assessed. This scenario has the longest modelling horizon, up to 2050, with projections starting in 2015 using a 5-year-step, while assuming that the legally binding targets of the 2020 Climate and Energy Package are achieved and that the EU and Member State policies agreed until December 2015[4] are implemented (Capros et al., 2016).

The seven mid-term scenarios, called EUCO, have been developed based on the REF2016 policy framework, but include additional policies and incentives across all Member States in order to assess the potential impact of achieving the EU's climate and energy targets by 2030 (E3MLab, 2019; E3MLab and IIASA, 2016):

---

[3] Trend projections that represent hypothetical, "what-if" scenarios, as a set of assumptions are made to address several unknowns ranging from population growth, macroeconomic and fossil fuel price developments, technology improvements, as well as the degree of policy implementation across the EU. On the contrary, forecasts would represent best estimates of future results and the most-likely scenario assumptions.
[4] Include the legally binding 2020 targets and EU legislation, the EU Emissions Trading System Directive (including the Market Stability Reserve), the Energy Performance of Buildings Directive, Regulations on eco-design and on $CO_2$ emission standards for cars and vans, as well as the revised F-gas (i.e. fluorinated greenhouse gases) Regulation. It also takes into consideration the Member States' explicit phase-out policies and ongoing projects at the time of modelling but does not incorporate the, at the time, politically agreed but not legally adopted, 2030 climate and energy targets.

- The EUCO27 and the EUCO30 target an at least 40% reduction in GHG emissions compared to 1990, with a 27% RES share in final energy use and a 27% and 30% energy efficiency target respectively (compared to 2007).
- The EUCO3030 and the three EUCO+ scenarios built on the EUCO30 to assess the potential impacts of more ambitious targets: EUCO3030 assesses a higher penetration of renewable energy of 30%, hence achieving a 43% GHG emissions reduction and the EUCO+33, EUCO+35 and EUCO+40 explore more ambitious energy efficiency targets of 33%, 35% and 40% respectively with a 28% RES share, thus reaching a GHG emissions reduction between 43% and 47%.
- The EUCO3232.5 is the most recent scenario that assess a raised 2030 target of energy efficiency to 32.5% and a RES share of 32%, as agreed in the 'Clean energy for all Europeans package', thus reaching one of the highest reductions in GHG emissions of 46%.

In addition, eight long-term, economy-wide scenarios, have been developed based on a "Baseline" policy framework that keeps the macro-economic projections, fuel price developments and pre-2015 Member States policies of the REF2016, but updates the technology pathways (De Vita et al., 2018), as well as projects the achievement of the 2030 energy and climate targets, thus are more comparable to the EUCO3232.5 pathway. These long-term scenarios target a GHG emissions reduction by 2050 ranging from -80% to -100% compared to 1990, while considering specific technology portfolios and assuming their deployment is intensified after 2030 (European Commission, 2018):

- Five scenarios target an 80% reduction of GHG emissions by 2050, of which **Electrification (ELEC), Hydrogen (H2) and E-fuels (P2X)** focus on switching fossil fuels to zero/carbon-neutral energy resources, whereas the **Energy Efficiency (EE) and** the Circular Economy (CIRC) explore more ambitious energy savings measures and a transition to a circular economy respectively.
- The **Combo** scenario, combines the above five scenarios without any predefined emissions reduction target for 2050, thus achieving close to 90% reduction in emissions by 2050.
- The **1.5TECH** and **1.5LIFE** scenarios target net-zero GHG emissions by 2050 (including land-use, land-use change and forestry), with the former relying more heavily on the intensification of the above technologies including the deployment of carbon-capture and storage technologies whereas the latter relies less on technological options and more on sustainable lifestyle changes e.g. increased climate awareness, carbon intensive diets, limiting growth of air transport demand etc.

As highlighted in the introduction, the above EC scenarios reflect technology-specific energy and climate targets and consider particular policy incentives for renewable energy sources (RES), all of which could influence the interplay of technologies in the power-mix and could in turn impact the EU power-system's cost-effective decarbonisation. To explore such implications, the role of nuclear power in the cost-effective decarbonisation of the EU power mix is explored based on data from a recent modelling study reflecting different levels of nuclear social acceptance (FTI Consulting, 2018).[5] The study covered three potential nuclear capacity scenarios for 2050, 150 GW (High), 103 GW (Medium) and 36 GW (Low), that reflect policy pathways of low towards higher social acceptance for nuclear power, considering as a base the current 2050 decarbonisation targets and the 'Clean energy for all Europeans package' 2030 targets of at least 40% and 85% reduction in GHG emissions by 2030 and 2050 respectively, compared to 1990 (similar to the EUCO33 scenario). This analysis focuses on assessing the potential medium-term and long-term (by 2050) impacts under three dimensions of energy security of supply, emissions reduction, and decarbonisation costs that are associated with the two extreme 2050 scenarios as follows:

a) **the low nuclear capacity scenario** represents a case where most of the existing plants close without further extensions and new projects fail to conclude, resulting in 36 GW of nuclear installed capacity by 2050;
b) **high nuclear capacity scenario**, models long-term operation extensions and building new plants in line with current advanced projects, as well as considers the commissioning of a number of additional plants (including about 5GW of SMR and <1GW of Gen-IV) that replace thermal baseload power plants, resulting in 150 GW of nuclear installed capacity by 2050.

---

[5] Assumes technology improvements based on the EC reference assumptions on electricity costs and performances, capital and operational expenditure based on latest data from EC and E3M, fuel commodity prices from IEA World Energy Outlook 2017, the $CO_2$ EUA to converge to EUCO33 by 2025 and EUCO30 by 2030/5, as well as outlook for new and existing Interconnections from the ENTSOE TYNDO 2018 Outlook.

## 3  Analysis of the EU Decarbonisation and Energy Transformation Trends

**Emissions, Energy Consumption and Import Dependency in the EU**

As illustrated in Figure 1, in contrast with worldwide trends, the EU economy has already started to modernise and transform, having experience a growth of over 60% since 1990, while at the same time reducing its greenhouse gas (GHG) emissions by 23%. However, definite conclusions cannot be drawn based on the latest available data, since the strongest reduction in emissions coincided with the period of 2007-2014 of the global financial and eurozone crises and from 2014 annual emissions have remained relatively constant. Furthermore, decoupling energy use and economic growth has shown to be a greater challenge, as annual energy consumption has mostly displayed an upward trend from 1990 onwards, with exception again the period coinciding with the financial crises. In addition, detailed look at the four sectors responsible for over 95% of EU's final energy use (Figure 2) reveals that compared to the 1990 levels, only industry has reduced its final energy use (-23%), while both transport and services have 1.3 times higher overall energy use compared to 1990 and only in households it increased marginally (+3.4%). Again, as of 2014, energy consumption has been experiencing again an upward trend in all sectors, except the residential one.

Figure 3 clearly shows that EU's import dependency has been mostly rising and the EU turned into a net-energy importer since the early 2000s, with a dependency reaching nearly 56 % in 2017. This trend is linked to the increase in natural gas dependency, as net imports of natural gas have more than doubled since 1990 and consumption increased by 32%.[6]

---

[6] Taking into consideration that both oil and petroleum as well as solid fossil fuel net-imports remained at close to 1990 levels but their consumption dropped.

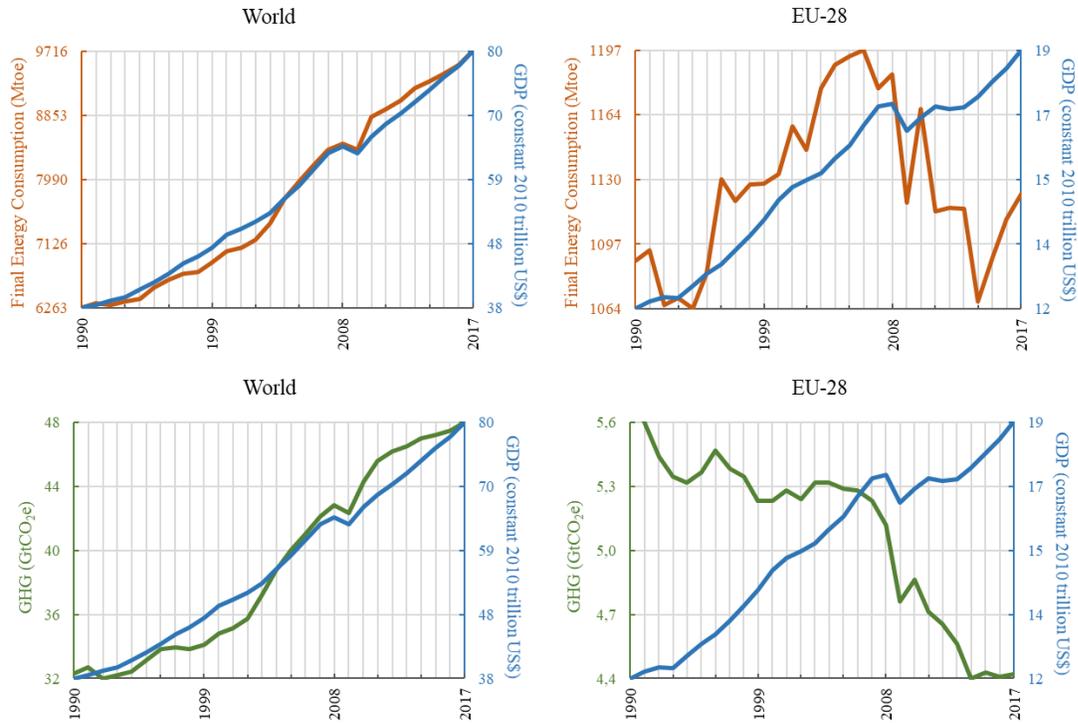

Figure 1: The global and EU28 annual change of energy consumption and GHG emissions compared to GDP growth for the period of 1990 to 2017 [based on data for GDP (World Bank, 2020), for emissions (Gütschow, et al., 2019) and for energy consumption (Eurostat, 2020; International Energy Agency, Organisation for Economic Co-operation and Development, 2020)]

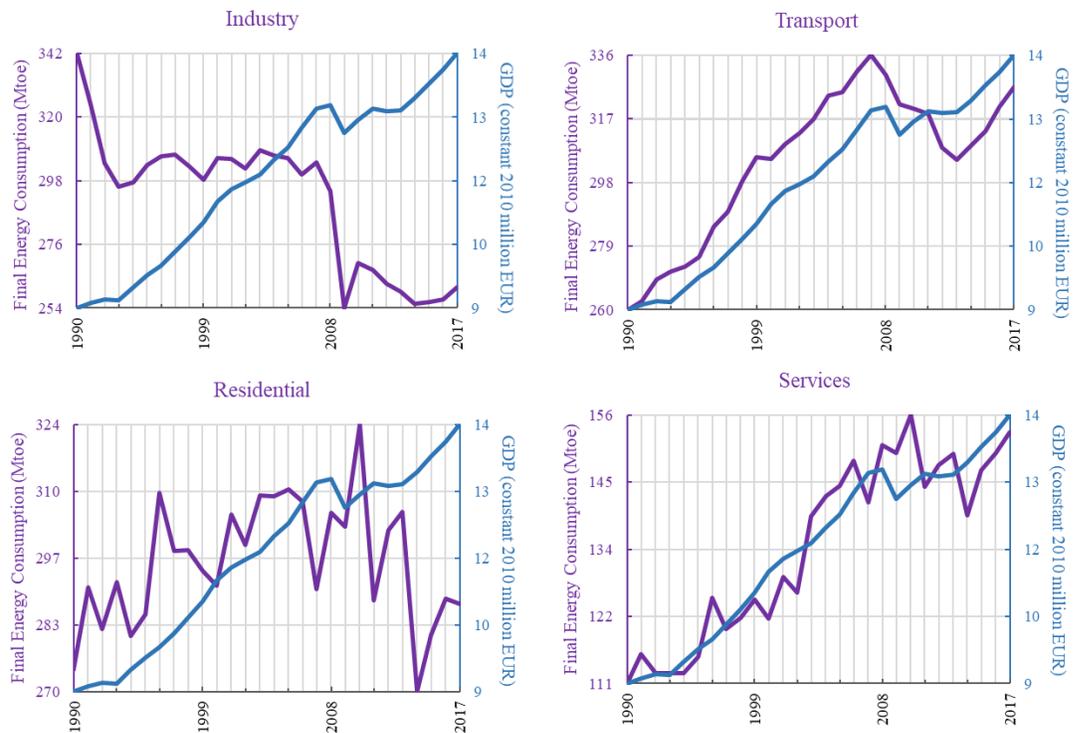

Figure 2: The EU28 annual change in final energy consumption and GDP by sector between 1990 and 2017 [based on data for GDP (World Bank, 2020), and for energy consumption (European Commission, Directorate-General for Energy, 2020)]

**Electrification and the Role of Nuclear**

The shares of electricity, natural gas and renewable heat & biofuels in the final energy use have increased compared to 1990, whereas those of fossil fuels decreased (Figure 4). Electricity is now the second most important energy fuel, together with natural gas, and it has also experienced the second-highest growth compared to 1990 (after renewable heat & biofuels), as a result of its increased use in industry and buildings (Figure 5), indicating the strong electrification of these two sectors. The industry sector was also the only sector that experienced a large drop in GHG emissions compared to 1990[7] (Figure 6), yet it is still the largest contributor of GHGs emissions (50%), and electricity and heat production activities retain by far the largest share, which has even slightly increased.

Interestingly, decarbonisation of electricity has been slower than the decarbonisation of manufacturing and construction and of fuel and petroleum refinement, despite electricity being increasingly generated by low-carbon technologies, with shares up to 56% in 2017, compared to 43% in 1990 (Figure 7). Nuclear technologies remain the most important source of low-carbon electricity, contributing almost half of it (45%) and accounting for only 11% of the total installed power capacity. The remaining half of low-carbon electricity is generated by RES which however add-up to four times the installed capacity of nuclear power. In addition, looking at the overall trends between 1990 and 2017, the shares of installed capacity of low-carbon generation technologies increased by 136%, owed, almost entirely, to new wind and solar power capacities (Table 1: Comparison of the EU28 gross electricity generation, installed generation capacity and operating hours by resource and technology between 1990 and 2017 [estimated based on data from (European Commission, Directorate-General for Energy, 2020)]

|  | Gross Electricity Generation (GWh) | | Installed Electricty Capacity (MW) | | Operating hours | |
|---|---|---|---|---|---|---|
|  | 1990 | 2017 | 1990 | 2017 | 1990 | 2017 |
| **Combustible fuels** | 1467060 | 1433590 | 331479 | 461069 | 4426 | 3109 |
| **Nuclear** | 794860 | 829720 | 121770 | 120885 | 6528 | 6864 |
| **Hydro** | 308710 | 331240 | 125327 | 155256 | 2463 | 2134 |
| **Wind** | 780 | 361940 | 454 | 168514 | 1718 | 2148 |
| **Solar** | 20 | 119400 | 11 | 109001 | 1818 | 1095 |
| **Geothermal & other RES (tidal)** | 3730 | 7250 | 739 | 1091 | 5047 | 6648 |
| **Biomass & alter. fuels (hydrogen)** | 19620 | 208860 | 4674 | 42231 | 4198 | 4946 |

**Electricity Costs and Prices**

According to published figures (European Commission, 2016b), electricity has held on average the highest share in energy-related household expenditures, which has been roughly half of the energy bill of an average EU household (European Commission, 2019b). Wholesale electricity prices, have displayed significant volatility over the last decade (e.g. 30 EUR'17/MWh in spring 2016 and 60 EUR'17/MWh in August 2018). These have been attributed to seasonality (i.e. wintertime vs. summertime) and variations in the generation mix, and as a result also to electricity production subsidies that influenced the electricity

---

[7] Referring to the GHG emissions resulting from the combustion of fossil fuels, which comprise ¾ of EU's GHG emissions. They include the GHG emissions from international aviation but exclude those from land use, land-use change, and forestry (LULUCF), as well as from international maritime

generation mix, since without these subsidies less RES capacity would have been installed in many EU countries over the past 10 years (European Commission, 2019d). On the other hand, and with exception large industrial consumers, retail prices grew by almost 20% for households and medium-sized industries between 2008 and 2017 (from 166 and 87 to 195 and 103 EUR'17/MWh respectively) (European Commission, 2019b). Similarly, these were heavily impacted by policy support costs and fiscal instruments, which changed significantly the price composition since 2008, in particular with respect to the share of taxes that almost doubled. The main driver for this increase in taxes was the support towards RES and combined-heat and power (CHP) generation, which, depending on the consumer's status, represented about 30-64% of the taxes and 12-24% of the average total electricity price, whereas the support for the nuclear sector was less than 1% of the price. Furthermore, the annual energy-related subsidies increased by 12% between 2008-2016, of which half were dedicated to support RES (with a threefold increase, reaching 76 billion EUR'17), 1/3 to fossil fuels and just 3% to nuclear (among the lowest at approximately 5 billion EUR'17); the remaining were dedicated to electricity, energy efficiency measures and heating and cooling (Rademaekers et al., 2018).

), yet the generation of low-carbon electricity increased by just 65%. This is expected as power generation by variable and intermittent RES (i.e. solar and wind) is highly depended on climate-related conditions (i.e. availability of wind and sun). These low-carbon technologies, together with hydropower, operated 1/3 to 1/6 of the time compared to nuclear power and the remaining time were idle.[8] Overall these trends translate to a smaller production of decarbonised electricity per area (i.e. the installed capacity representing a proxy of the area required) and to a higher underutilisation of facilities and hence investments.

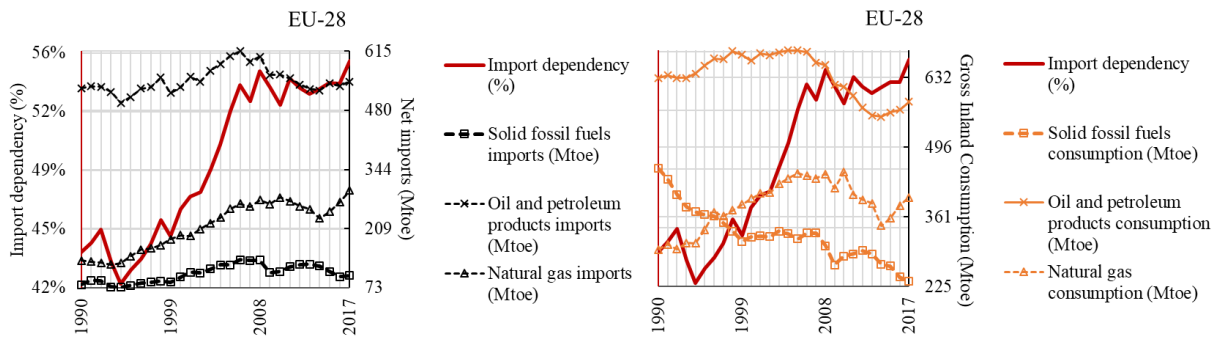

Figure 3: The EU28 annual change in import dependency compared to the net imports (left) and gross inland consumption (right) by energy source between 1990 and 2017 [estimated based on data from (European Commission, Directorate-General for Energy, 2020)]

---

[8] It should be noted that the operation of dispatchable power plants, and hence their capacity factors (also called load factors), can be further constrained by market forces, as the distribution system operator might give priority to generating installations using renewable energy sources, waste or CHP, according to Regulation (EU) 2019/943 (European Parliament and Council of the European Union, 2019). Among the typical dispatchable generating plants are natural gas and hydro, while nuclear and coal-fired plants are theoretically dispatchable yet designed as baseload plants.

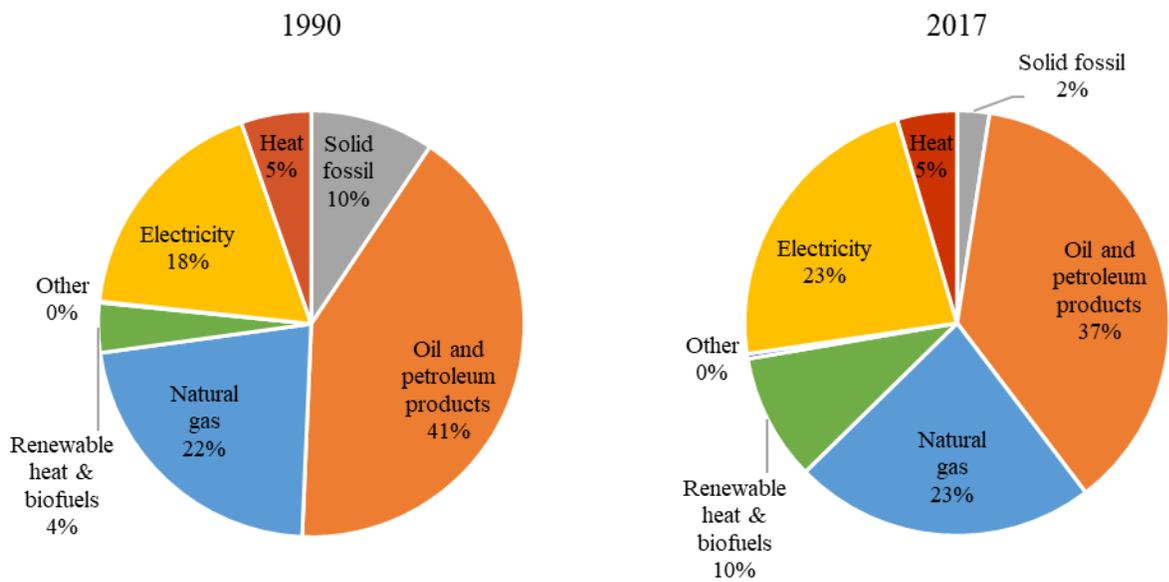

Figure 4: Comparison of the 1990 and 2017 shares (%) of energy fuels (i.e. resources and carriers) in the total final energy consumption of the EU28 [estimated based on data from (European Commission, Directorate-General for Energy, 2020)]

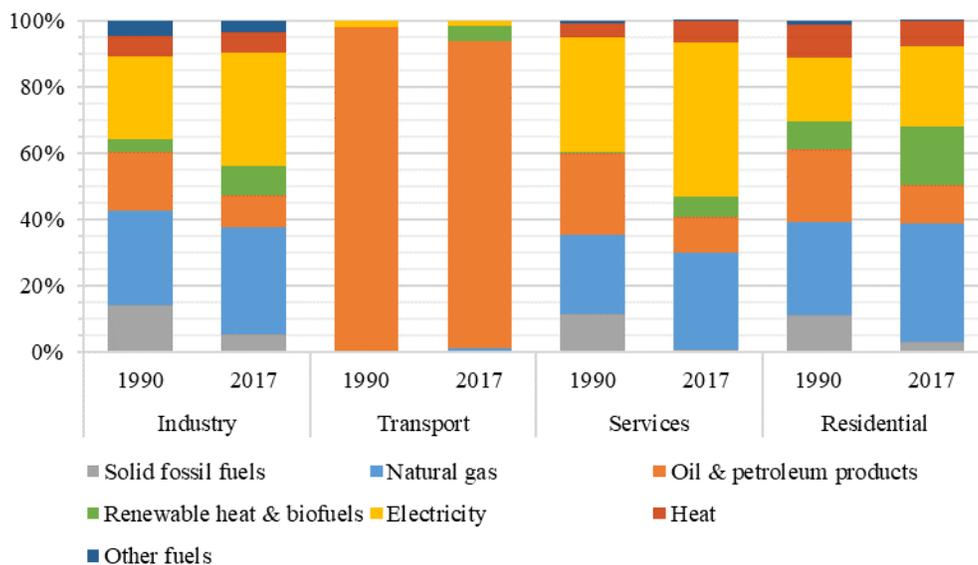

Figure 5: Comparison of the EU28 sectoral shares (%) of final energy consumption between 1990 and 2017 by fuel type [based on data from (Eurostat, 2020)]

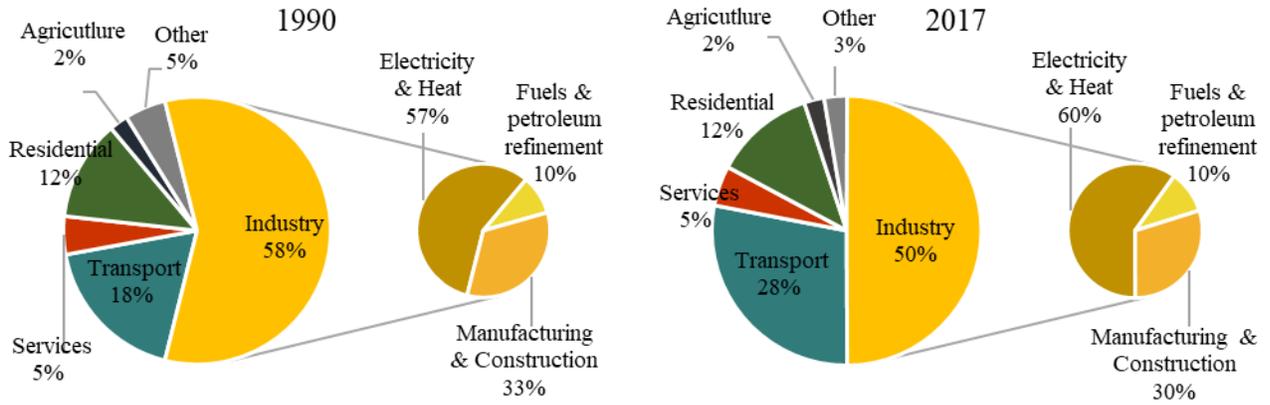

Figure 6: Comparison of the 1990 and 2017 sectoral shares (%) of GHG emissions in the EU28 [estimated based on data from (European Commission, Directorate-General for Energy, 2020)]

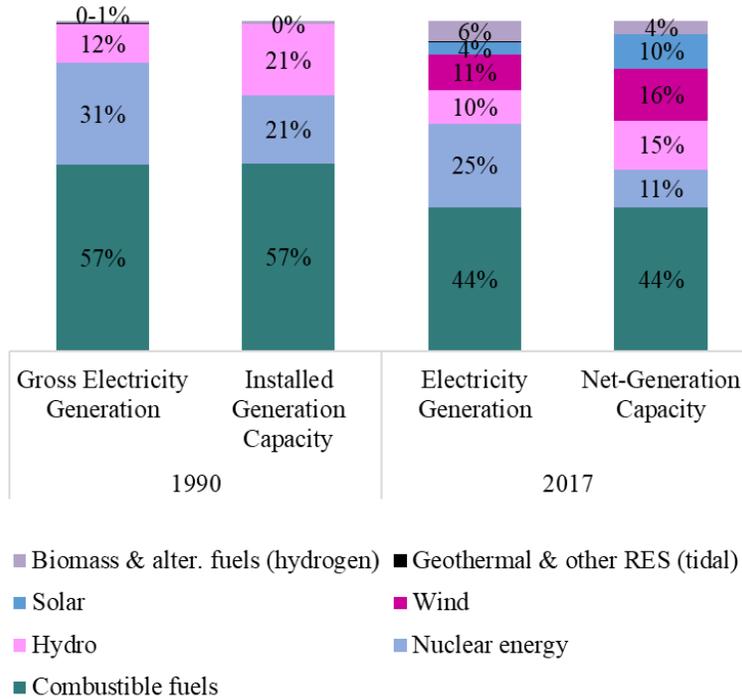

Figure 7: Comparison of the EU28 shares of gross electricity generation and installed generation capacity by source between 1990 and 2017 [estimated based on data from (European Commission, Directorate-General for Energy, 2020)]

Table 1: Comparison of the EU28 gross electricity generation, installed generation capacity and operating hours by resource and technology between 1990 and 2017 [estimated based on data from (European Commission, Directorate-General for Energy, 2020)]

|  | Gross Electricity Generation (GWh) | | Installed Electricty Capacity (MW) | | Operating hours | |
| --- | --- | --- | --- | --- | --- | --- |
|  | 1990 | 2017 | 1990 | 2017 | 1990 | 2017 |
| **Combustible fuels** | 1467060 | 1433590 | 331479 | 461069 | 4426 | 3109 |
| **Nuclear** | 794860 | 829720 | 121770 | 120885 | 6528 | 6864 |
| **Hydro** | 308710 | 331240 | 125327 | 155256 | 2463 | 2134 |
| **Wind** | 780 | 361940 | 454 | 168514 | 1718 | 2148 |
| **Solar** | 20 | 119400 | 11 | 109001 | 1818 | 1095 |
| **Geothermal & other RES (tidal)** | 3730 | 7250 | 739 | 1091 | 5047 | 6648 |
| **Biomass & alter. fuels (hydrogen)** | 19620 | 208860 | 4674 | 42231 | 4198 | 4946 |

**Electricity Costs and Prices**

According to published figures (European Commission, 2016b), electricity has held on average the highest share in energy-related household expenditures,[9] which has been roughly half of the energy bill of an average EU household (European Commission, 2019b).[10] Wholesale electricity prices, have displayed significant volatility over the last decade (e.g. 30 EUR'17/MWh in spring 2016 and 60 EUR'17/MWh in August 2018).[11] These have been attributed to seasonality (i.e. wintertime vs. summertime) and variations in the generation mix,[12] and as a result also to electricity production subsidies that influenced the electricity generation mix, since without these subsidies less RES capacity would have been installed in many EU countries over the past 10 years (European Commission, 2019d). On the other hand, and with exception large industrial consumers, retail prices grew by almost 20% for households and medium-sized industries between 2008 and 2017 (from 166 and 87 to 195 and 103 EUR'17/MWh respectively) (European Commission, 2019b).[13] Similarly, these were heavily impacted by policy support costs and fiscal instruments, which changed significantly the price composition since 2008, in particular with respect to the share of taxes that almost doubled. The main driver for this increase in taxes was the support towards RES and combined-heat and power (CHP) generation[14], which, depending on the consumer's status, represented about 30-64% of the taxes and 12-24% of the average total electricity price, whereas the support for the nuclear sector was less than 1% of the price. Furthermore, the annual energy-related subsidies increased by 12% between 2008-2016, of which half were dedicated to support

---

[9] In comparison with gas, solid and liquid fuels, and heating but excluding transport expenditures, based on.
[10] Since these are average values, it is important to note that large differences are observed across the EU on both the absolute expenditures, the share of energy in the total expenditure and the role of different household energy products.
[11] All prices are expressed in constant EUR of the year 2017.
[12] Since the marginal costs of generation technologies have been dominated by the higher coal and natural gas prices. Marginal costs refer to the incremental cost of producing one more unit of electricity and reflect the change in the total cost.
[13] Large industry consumers pay lower taxes due to partial exemptions from levies according to the Energy and Environment State Aid Guidelines (EEAG).
[14] This considers only policy support costs that directly impact retail prices and not every tax sub-component existing in each Member State.

RES (with a threefold increase, reaching 76 billion EUR'17), 1/3 to fossil fuels and just 3% to nuclear (among the lowest at approximately 5 billion EUR'17); the remaining were dedicated to electricity, energy efficiency measures and heating and cooling (Rademaekers et al., 2018).

# 4 Comparison of Medium- and Long-term Decarbonisation Scenarios and Possible Socio-economic Implications

## 4.1 The 2030 Scenarios and the Role of Nuclear

**Energy Security of Supply and Emissions Reduction**

All EC mid-term scenarios model a drop in the gross energy demand, with RES playing a more significant role driven by the 2030 targets, while the importance of natural gas and nuclear is projected to remain relatively stable and of fossil fuels to decrease (Figure 8). As expected, a reduction in the gross energy demand would reduce net energy imports, yet the EU is projected to remain a net importer until 2030, with import dependency above 50%, even in the EUCO3232.5 scenario. This can be attributed to the projected increase in natural gas imports to 1.2-2 times above the 1990 levels, accounting for 1/4 to 1/3 of the total net imports by 2030, depending on the scenario.

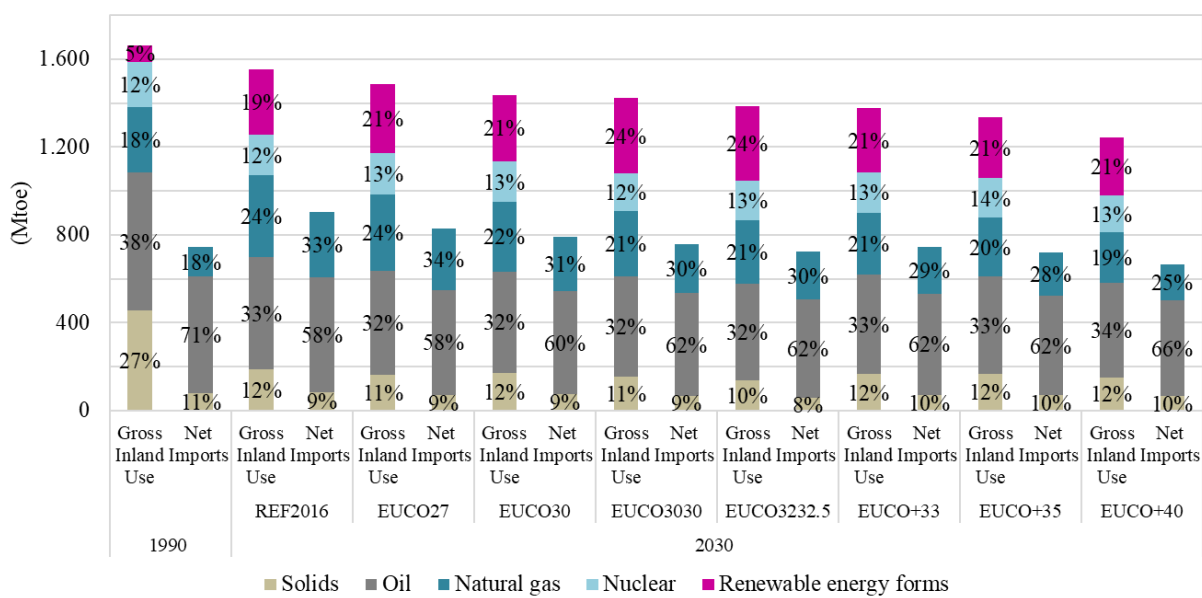

Figure 8: Mid-term projections (in 2030) of the gross inland consumption (excluding non-energy uses of energy resources and the consumption of the energy sector itself) and the net imports compared to 1990 in the EU28 [estimated based on data from (E3MLab, 2019; E3MLab and IIASA, 2016; European Commission, Directorate-General for Energy, 2020)]

Electricity is the only energy fuel projected to further increase its share in the final energy use, remaining the second most important energy fuel (Figure 9 compared to Figure 4). Focusing on the EUCO3232.5 scenario, which represents more closely the current EU current policy framework, electricity is expected to continue decarbonising: low-carbon generation would increase by 30% by 2030 compared to 2017, yet low-carbon power capacity would increase by 125% (Figure 10 and Figure 7). The role of nuclear power in decarbonising electricity would remain significant, being its second largest contributor despite a projected reduction of nearly 10% in its installed capacity compared to 2017. It can be estimated from the figure, that by 2030 the projected installed capacity for RES would be 4 times that of nuclear, to generate an equivalent amount of low-carbon electricity. This observation is in line with the historical trends analysed in Section 2 and no significant change in the operating hours has been noticed since 2017.

Moreover, despite the significant changes in the power-generation mix that target a large drop in emissions by 2030, electricity & heat is projected to be the second largest emitter, accounting for about 27% of all energy-related $CO_2$ emissions (Figure 11).

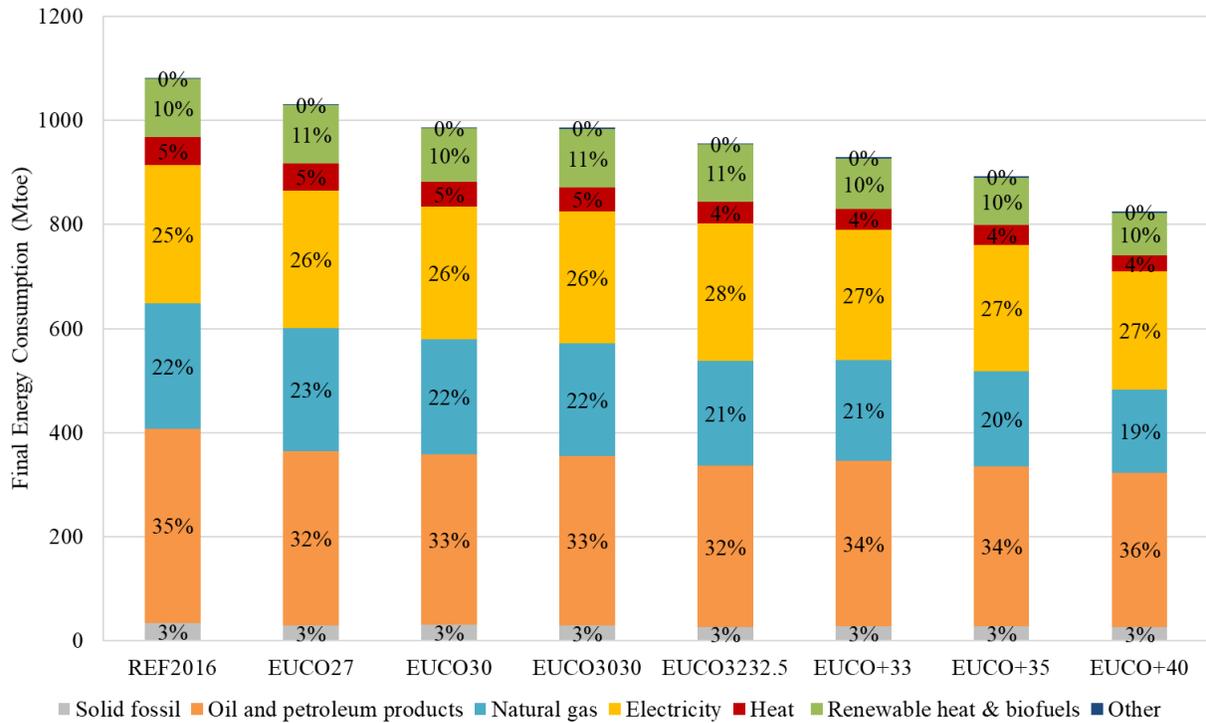

Figure 9: Mid-term projections (in 2030) of the shares of energy fuels (i.e. resources and carriers) in the final energy use in the EU28 [based on data from (Capros et al., 2016; E3MLab, 2019; E3MLab and IIASA, 2016)]

Analysing the potential medium-term implications of a policy pathway of low social acceptance for nuclear power as compared to a pathway of high social acceptance (FTI Consulting, 2018), indicates a required increase in thermal installed capacity of 27 GW in order to compensate for the loss in nuclear installed capacity, of which about 2/3 would be for new infrastructure and the remaining extensions to existing carbon-intensive units. This change in the power generation mix is projected to bring an increase in fossil-fuel consumption of 5'112 TWh, (53% natural gas and 47% coal) and result in a missed opportunity to achieve an additional reduction in $CO_2$ emissions of 1971 Mt by 2030. Such differences in infrastructure investments would risk carbon lock-in, stranded assets and heighten EU's energy dependence for the coming decades, whereas reaching a higher level of emissions in the mid-term, would not allow the EU to achieved its full potential by 2030, and at the same time could hinder EU's long-term decarbonisation transition before the roll-out of RES and storage, putting an upward pressure on EU ETS prices and end-customer costs.

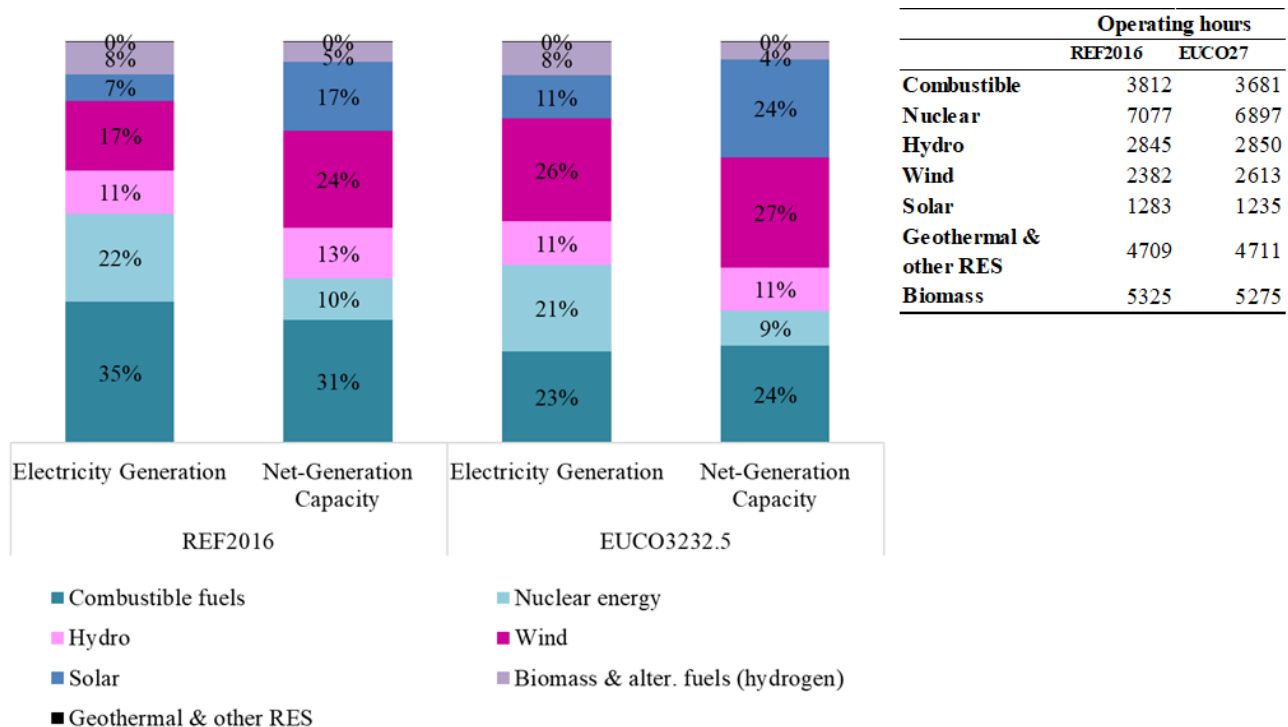

Figure 10: Mid-term projections (in 2030) of the electricity generation and net-generation capacity by technology in the EU28 [estimated based on data from (Capros et al., 2016; E3MLab, 2019)]

**Decarbonisation Costs**

In terms of electricity production costs, all scenarios project them to peak in the 2020s[15] and subsequently slightly drop in the 2030s (Table 2).[16] Yet, electricity production costs are expected to display a much higher daily variability in the future, with average values continuing to be driven by fossil fuel plants, but also increasingly by biomass-fired plants. The price of electricity is projected to keep rising, although at a smaller pace that observed by the historical trends (Section 2) and is expected to increase by another 10% compared to 2020, with total energy-related costs increasing by 22% in the same period. Compared to all mid-term scenarios, the EUCO3232.5 scenario requires the largest amount of investments by 2030, with solar PV and wind power capacities projected to continue dominating new investments, accounting for roughly 90% of new installed capacity between 2015 and 2030 (Figure 10). The scenarios project a significant decrease in fossil-fuel-fired power capacities (-92%) and a nearly 10% drop in nuclear and natural gas power capacities by 2030, since it is projected that increases in electricity needs would be satisfied by achieving the RES 2030 targets. This is expected as the current gas-fired power plant fleet is relatively young and would still be operational by 2030 (European Commission, 2019b).

---

[15] The analysis is not directly comparable with the historical figures discussed in Section 3 as the modelling focused on trends and relative changes and only average values were included in the technical reports in constant EUR of the year 2013. In addition, the costs for storage and additional interconnections are not accounted for in the projections.

[16] Projections are similar to those of the Central Scenario developed by the Commission's JRC energy model POTEnCIA (Mantzos et al., 2019).

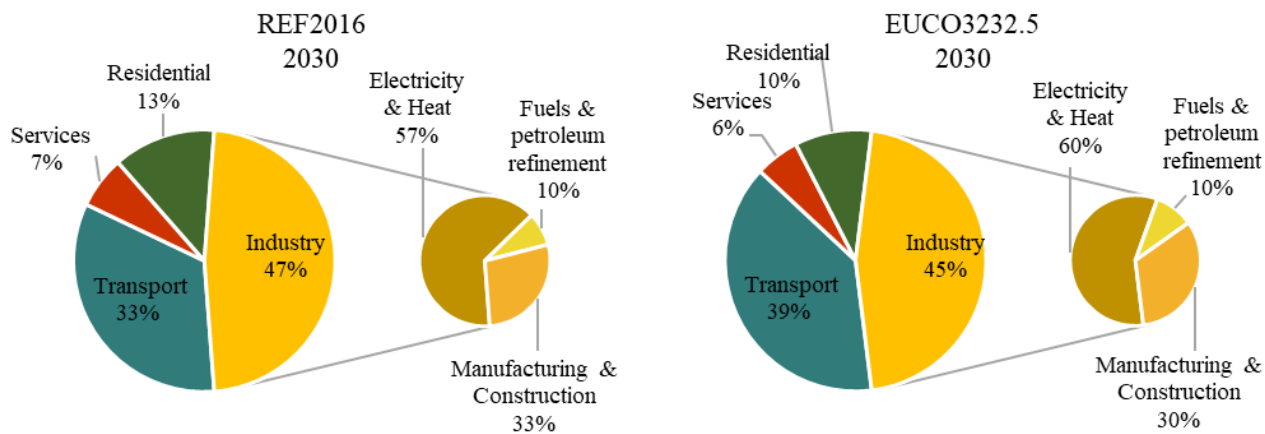

Figure 11: Mid-term projections (in 2030) of the energy related $CO_2$ emissions by sector in the EU28 [estimated based on data from (Capros et al., 2016; E3MLab, 2019)]

Table 2: Mid-term projections (in 2030) of the electricity and energy-related costs in the EU28 [based on data from (Capros et al., 2016; E3MLab, 2019)]

|  | 2010 | | 2020 | | 2030 | | '10-'20 | | '20-'30 | |
|---|---|---|---|---|---|---|---|---|---|---|
|  | REF2016 | EUCO3232.5 | REF2016 | EUCO3232.5 | REF2016 | EUCO3232.5 | REF2016 | EUCO3232.5 | REF2016 | EUCO3232.5 |
| Average Cost of Gross Electricity Generation (€'13/MWh) | 65 | 65 | 94 | 93 | 91 | 89 | 3,8 | 3,7 | -0,4 | -0,5 |
| Average Price of Electricity in Final demand sectors (€'13/MWh) | 136 | 136 | 153 | 151 | 161 | 163 | 1,2 | 1,1 | 0,5 | 0,7 |
| Total energy-related costs* (in billion €13) | 1468 | 1467,9 | 1791 | 1782,9 | 2032 | 2166,6 | 2,0 | 2,0 | 1,3 | 2,0 |
| as % of GDP | 11,4 | 11,4 | 12,3 | 12,3 | 12,2 | 13,0 |  |  |  |  |

* Includes the capital, operating expenditures and efficiency investment costs, excluding costs related to auctioned emissions. Capital expenditures and energy efficiency investment costs have been discounted at a 10% rate.

Analysing the potential medium-term implications of an policy pathway of low social acceptance for nuclear power as compared to a pathway of high social acceptance (FTI Consulting, 2018), indicates that an early closure of NPPs closure and limited nuclear investments would trigger an additional increase in electricity prices of 20 EUR/MWh in the medium-term. This is associated with the increased use of gas- and coal-fired plants that set wholesale prices in the mid-term, whereas the losses in end-user savings[17] are estimated at 206 billion EUR'17 (undiscounted), because cheaper nuclear baseload is replaced by more expensive gas and coal generation. In terms of capital investments, although temporary gains are expected in the medium term due to the lack of commissioning and retrofitting of NPPs, these are projected to be outweighed in the long-term and will therefore be discussed in the subsection below[18].

---

[17] The end-user costs include all costs associate with providing electricity at the end point (electricity generation cost, generation capacity cost and subsidy costs), and are comparable to the total energy system costs.

**Potential Socio-economic Externalities**

Rough estimates of possible medium-term implications with respect to externalities indicate that an early closure of nuclear power plants could translate into a missed opportunity to create 204'000 high skilled job-years in the nuclear generation sector by 2030 in the EU, as nuclear power is the most job intensive technology in terms of direct employment per site and the second most direct job intensive technology (i.e. 0.5 Jobs/MW based on (NEA, 2018)). Moreover, by replacing nuclear power with less energy-dense power technologies, could result in land-use[19] losses of approximately 2852 km$^2$ by 2030, an area almost equivalent to the total area of Luxembourg. Finally, the increase in additional thermal generation in order to compensate for limiting low-carbon nuclear power generation, could exacerbate EU's air pollution by almost 10 Mt of SOx and NOx emissions, and 3230kt of PM by 2030.

Future investments beyond 2030 depend on the assumed lifetimes of the current power plants and as already mentioned, the majority of the current power plant fleet would still be operational by 2030. In the long-term, the influence of policy pathways on the power mix would become more evident, since a substantial part of the current power plants (about 1/3 of the capacity installed prior to 2005) would reach the end of their lifetime operation between 2030 and 2040. This means that a faster pace of phase-out policies with deeper decarbonisation and electrification, could substantially change investment needs, and different policy routes could have a varying-level of impact on the future technology deployment, while at the same time, most of the assumed techno-economic developments take effect, as scenarios presume storage and CCS technologies to become cost effective from 2040 onwards (Capros et al., 2016; Simoes et al., 2017). These changes are expected to trigger rapid investments that, depending on the policy pathways that would translate into a fundamentally different power mix by 2050.

---

[18] This applies also to flexible sources, i.e. batteries and long-term storage including P2G, which the models assume to be cost-effective after 2035-2040, as well as grid and balancing costs, which appear after 2030 (Capros et al., 2016).

[19] Land-use costs are difficult to be estimated and are approximated using the geographic footprint.

## 4.2 The 2050 Scenarios and the Role of Nuclear

**Energy Security of Supply and Emissions Reduction**

By 2050, gross inland energy use is projected to show large variations depending on the policy pathway followed and the decarbonisation ambition. Compared to the pathway based on existing policy measures (i.e. Baseline), on one hand energy use could be lower driven by efficiency developments, electrification, the circular economy and by an increased decarbonisation ambition relying on sustainable lifestyle changes (i.e. EE, ELEC, CIRC and 1.5LIF). On the other hand it could also be at similar levels (i.e. H2 and COMBO) or even increase to meet the growing needs for the production of hydrogen and e-fuels (i.e. P2X and 1.5TECH) (Figure 12). The Baseline projects the EU to remain a net-energy importer in 2050, whereas the EU's import dependency is expected to improve significantly in all other scenarios, with renewables followed by nuclear becoming the prevalent energy resources. Moreover, the higher the decarbonisation ambition (i.e. COMBO, 1.5TECH and 1.5LIFE) the higher increase is projected for the consumption of renewables and nuclear energy and the lowest for natural gas and other fossil fuels (compared to the EUCO3232.5).

All scenarios project electricity to become the dominant energy fuel by 2050, representing half of the final energy demand (Figure 13), however, individual scenario projections on the shares of energy technologies in electricity generation are not provided (European Commission, 2018), and so no analysis could be performed on the differences in the power-mix and decarbonisation ambition. At the same time, discrepancies were found between the reported values for nuclear electricity generation and its share in the final energy use that require further clarification before including them in such an analysis.[20] Nevertheless, the power sector is projected to have the most potential, displaying the steepest reduction in emissions across all 2050 EC pathways, which is mostly achieved through improved energy efficiency, a circular economy and deepening of electrification (i.e. CIRC, EE, ELEC).

All EC 2050 policy scenarios project a striking increase in installed power capacity compared to 2030, which more than doubles in the scenarios achieving 100% reduction in emissions (Figure 14). Similar to the mid-term projections, this trend is linked to projected increase in the installed capacities of solar and wind power compared to nuclear, that is expected to require massive investments in electricity infrastructure and storage technologies, especially in deep decarbonisation pathways and despite adopting policies to support energy efficiency, circular economy and sustainable lifestyle choices (i.e. COMBO, 1.5TECH and 1.5LIFE) (Figure 15). Specifically, and regardless of the EC policy pathway, solar and wind power capacities are projected to continue dominating new investments (77-88%), compared to new nuclear power capacities (5-7%),[16] whereas the highest investment needs in conventional storage (i.e. pumped hydro and batteries) would take place in the scenarios assuming less significant development of e-fuels (i.e. EE, CIRC and ELEC).

---

[20] Specifically, only an averaged decarbonisation scenario is provided in the supportive data (i.e. across all long-term scenarios), where nuclear power shares drop to 13% of electricity production, which is much lower than its projected share in the final energy consumption (i.e. 17-19% in Figure 13). On the contrary, the average share of RES in electricity generation is projected at 83% of which 69% is wind and solar, and the average fossil-fuel shares is at 4% (Figure 23 in (European Commission, 2018))

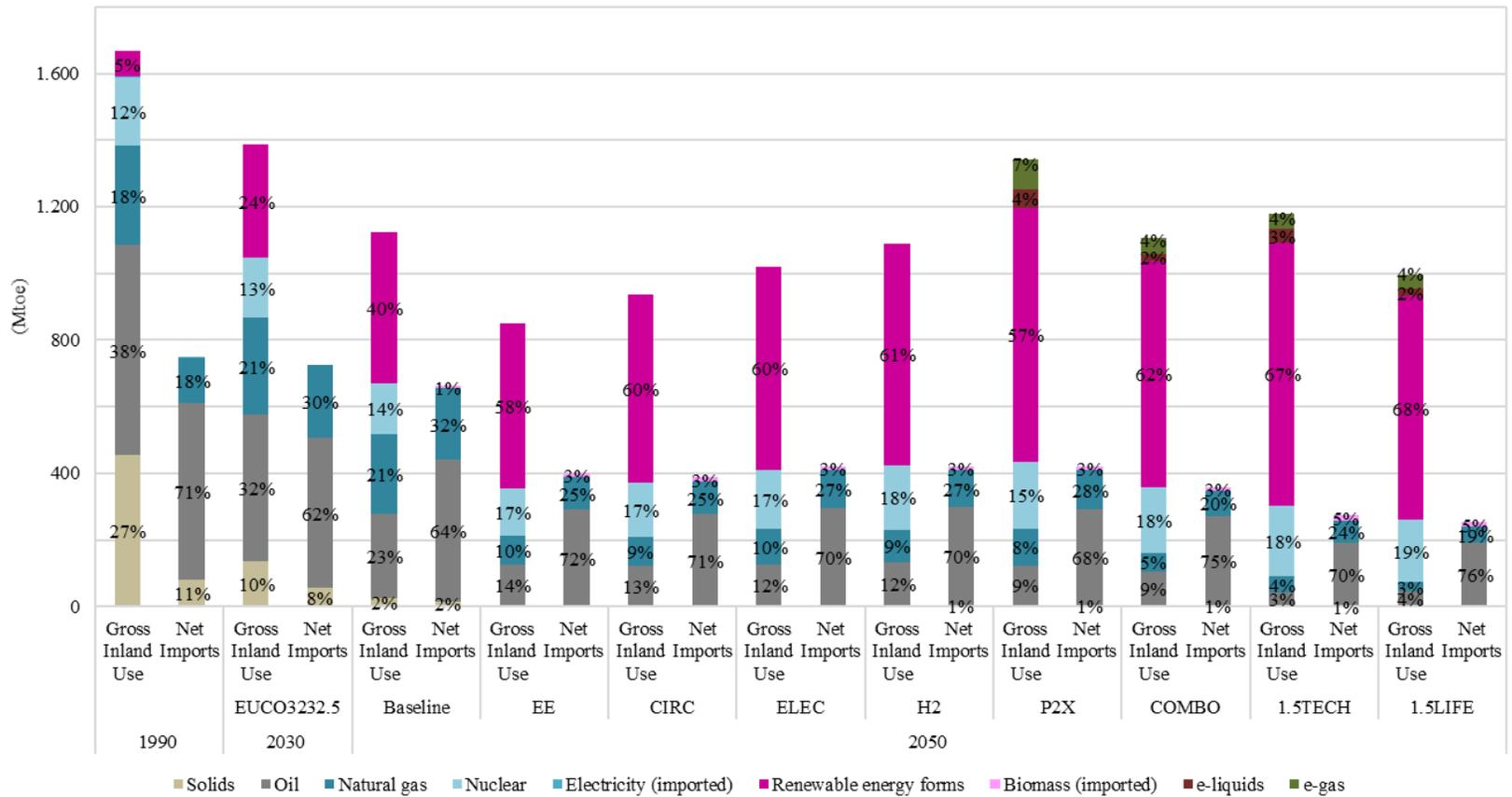

Figure 12: Long-term projections (in 2050) of the gross inland consumption (excluding non-energy uses of energy fuels and the consumption of the energy sector itself) and the net imports compared to 1990 in the EU28 [estimated based on data from (European Commission, 2019c; European Commission, Directorate-General for Energy, 2020)]

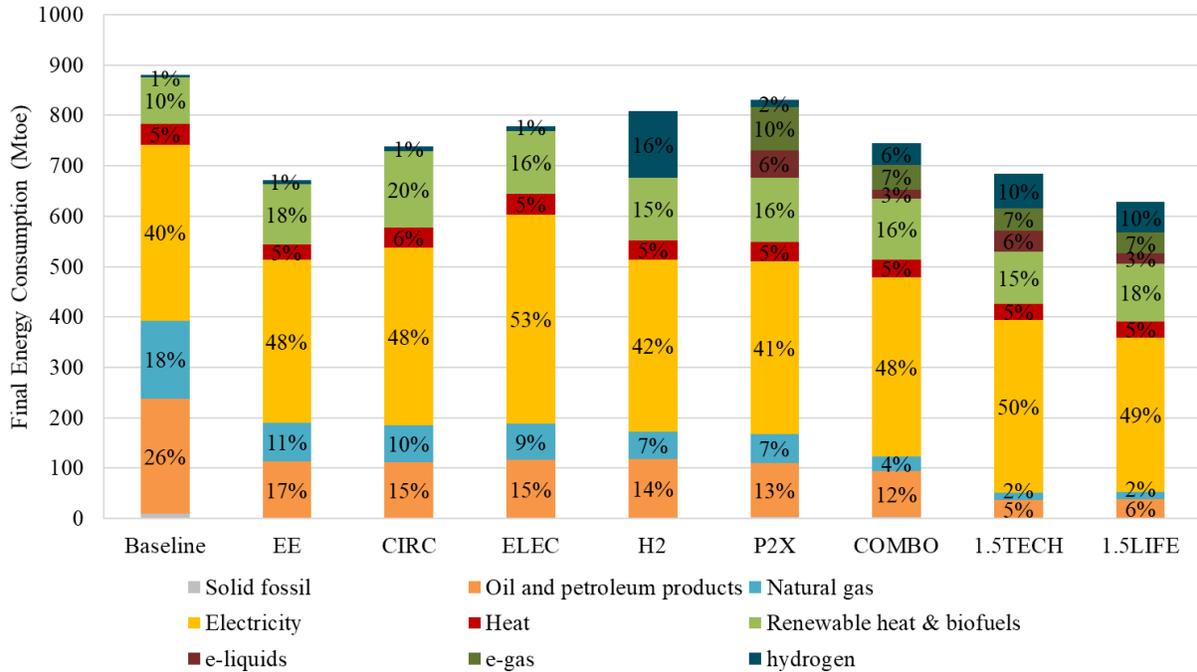

Figure 13: Long-term projections (in 2050) of the shares of energy fuels (i.e. resources and carriers) in the final energy use in the EU28 [estimated based on data from (European Commission, 2019c)]

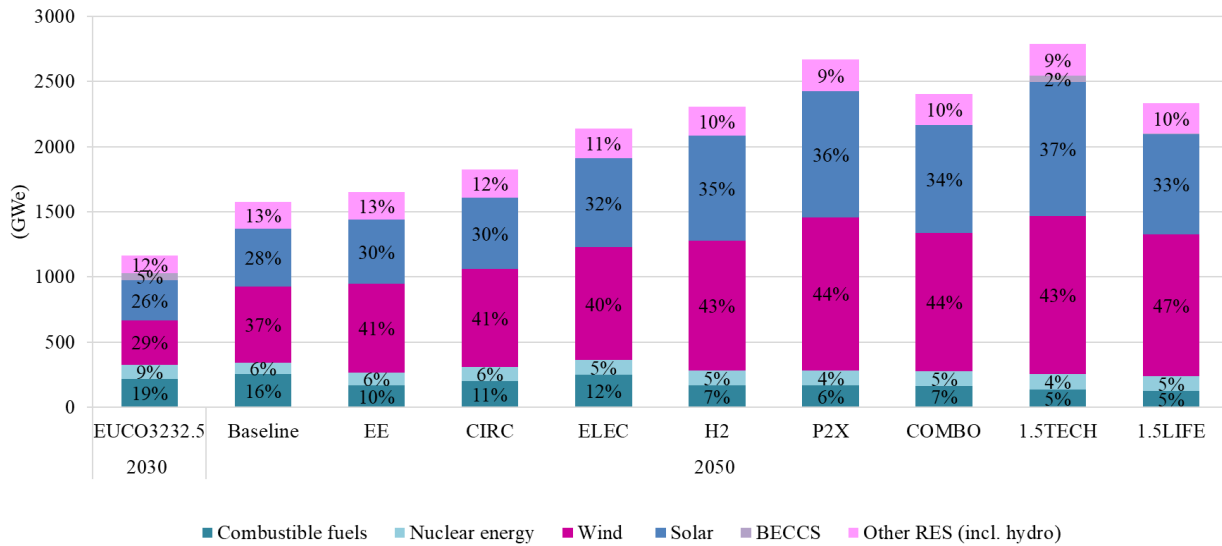

Figure 14: Long-term projections (in 2050) of the installed generation capacity by technology in the EU28 [estimated based on data from (E3MLab, 2019; European Commission, 2019c)]

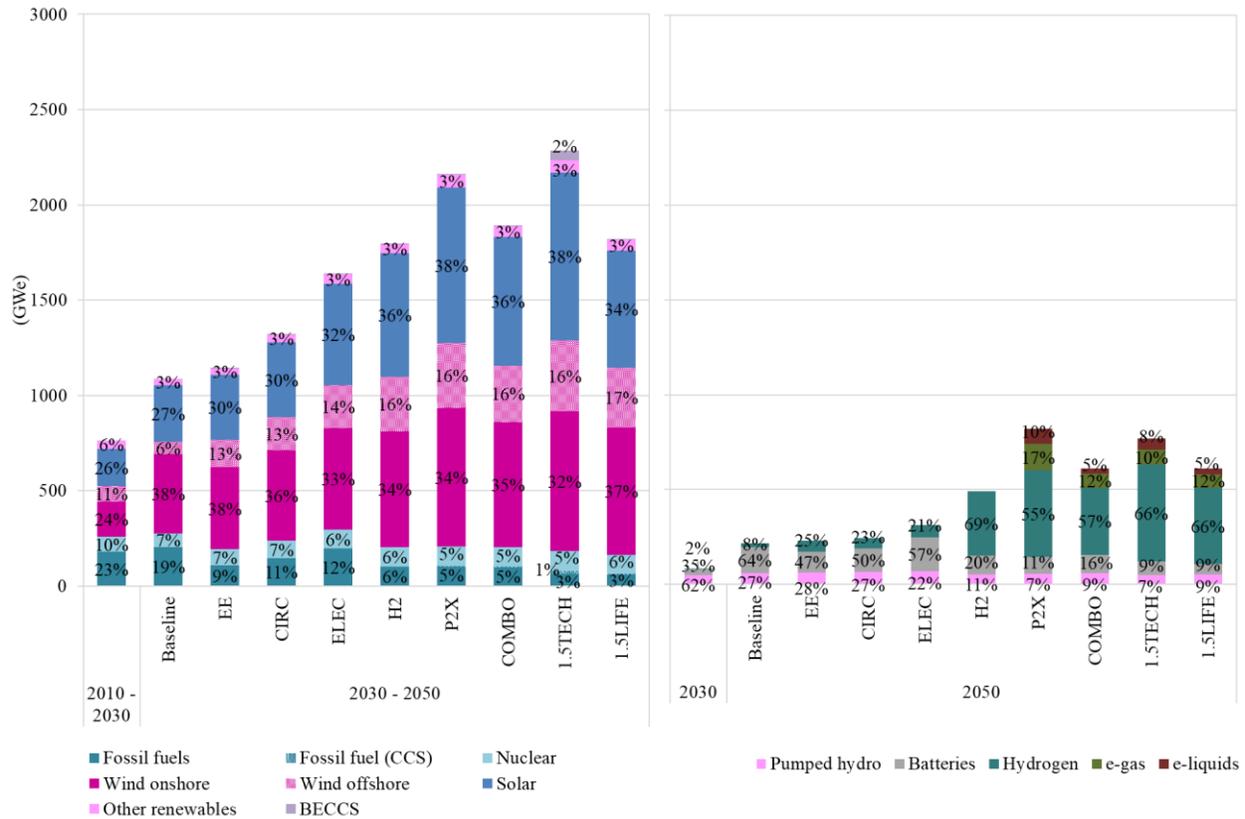

Figure 15: Projections of the (left) new, total, generation capacities by source, including new constructions, life-time extensions, refurbishment and retrofitting, and (right) capacities in electricity storage technologies, needed between 2030 and 2050 in the EU28 [estimated based on data from (European Commission, 2019c)]

Analysing the potential long-term implications of a policy pathway of low social acceptance for nuclear power as compared to a pathway of high social acceptance (FTI Consulting, 2018) indicates that in order to compensate for the potential loss of 114 GW of nuclear capacity, aside from the 25 GW of additional thermal capacity already required by 2030, additional investments in alternative technologies equivalent to 533 GW between 2020 and 2050 would be required (about 415GW of RES and 93 GW in battery and power-to-gas storage). These trends could induce an additional curtailed energy by RES of about 66TWh (or 1% of their total power) and give rise to fossil-fuel power generation (+36 % in natural and +18% of coal) to ensure security of supply, of which 80% would be consumed by 2030. Moreover, the EU's power system would rely more heavily on yet-to-be-proven storage technologies (European Commission, 2019e), of which 36 GW would need to be operational already between 2035 and 2040, while at the same time risking a higher import dependency on battery materials and generating large volumes of end-of-life batteries that currently have an average 10-year lifetime (Bobba et al., 2018). As a result, and despite the increased RES penetration, these would translate into a missed opportunity to further reduce $CO_2$ emissions by 2270Mt that would bring an additional 17% cut in the power sector emissions by 2050 and support EU's path towards clean electrification. Moreover, 95% of the total emission-reduction potential occurs by 2030 (see Section 4.1), which is projected to be an important worldwide tipping point for limiting worldwide warming to 1.5°C (Intergovernmental Panel on Climate Change, 2018), hence justifies the urgency for early action to reach EU's fullest potential and strengthen its role as a global leader in climate action.

**Decarbonisation Costs**

Depending on the policy pathway, investment costs are projected to increase at a different pace, mostly owed to differences in the technology deployment and decarbonisation ambition (Figure 16): the 1.5TECH pathway would require the largest amount of investments, followed by P2X, COMBO and H2, with energy-related costs increasing by 16 to 30% from 2030. Electricity prices however, follow diverging trends compared to 2030, forming seemingly three policy scenario groups by 2050: a) electricity prices drop with a low decarbonisation ambition, as costs are mitigated by improved efficiency and circular economy measures and intensified electrification (i.e. BL, ELEC, CIRC EE), b) electricity prices increase moderately (12-19%), either with a high decarbonisation ambition that relies on sustainable lifestyle choices (i.e. 1.5LIFE) or with a low decarbonisation ambition that relies on specific technological options (H2, P2X, COMBO), or a), and c) electricity prices increase drastically (almost 50%) with a high decarbonisation ambition that relies extensively on specific technological options, including CCS (1.5TECH).

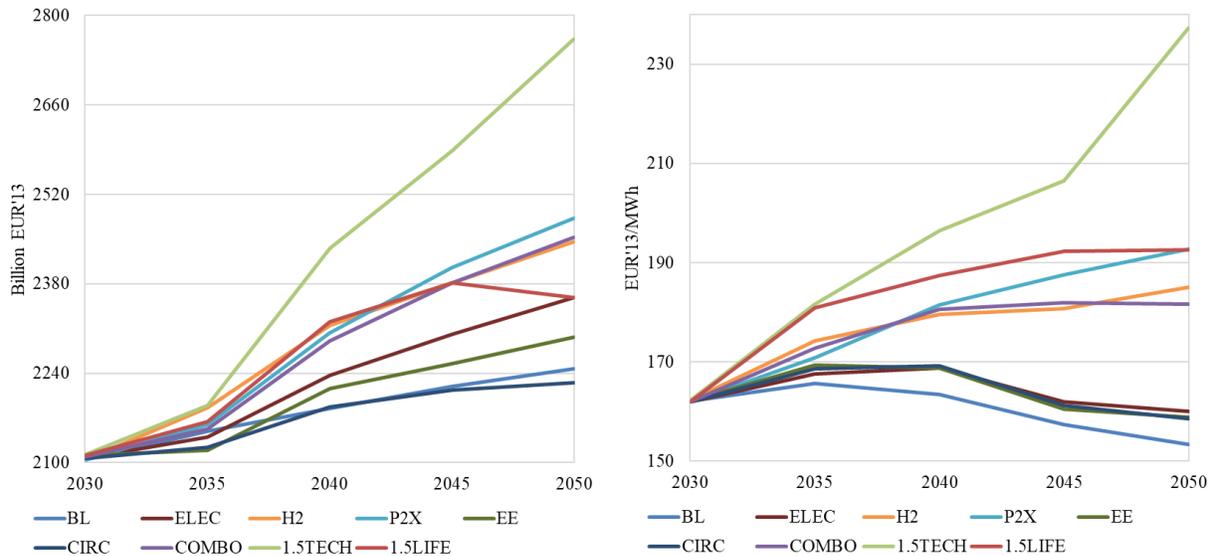

Figure 16: The evolution of the total (left) energy-related costs and (right) price of electricity by scenario up to 2050 in the EU28 [estimated based on data from (European Commission, 2019c)]

Analysing the potential long-term implications of a policy pathway of low social acceptance for nuclear power as compared to a pathway of high social acceptance (FTI Consulting, 2018) indicates that although power prices converge over time, price volatility would increase significantly due to the increased RES penetration. In addition, a peak in power-price losses is expected by 2030 (20 EUR/MWh) that could hamper the competitiveness of electricity against other energy fuels and slow down the electrification of sectors, such as transport. Furthermore, the overall losses for the end-users[17] are estimated at 350 billion EUR'17 (undiscounted), of which 90% would occur before 2035. Finally, although an early nuclear closure and limited investments, would save 75 billion EUR'17 by 2035, a doubling in investment losses is expected thereafter, which could result in an overall investment loss of 85 billion EUR'17 between 2020 and 2050. It is important to note that these economic figures were based on the assumption that the nuclear CAPEX would decrease by 37% by 2050, and although end-user costs were shown to be robust to changes in the CAPEX reduction, investment costs were highly sensitive (FTI Consulting, 2018), which further justifies the importance of recognising the contribution of nuclear power to decarbonisation and planning long-term financial support schemes to leverage the technological benefits of the nuclear power sector.

**Potential Socio-economic Externalities**

Rough estimates of possible medium-term implications with respect to externalities indicate that an early closure of nuclear power plants could translate into a missed opportunity to create 1 million high-skilled job-years in the EU nuclear power sector over 2020-2050. In addition early nuclear closure is expected to escalate the RES capacity and as a result both transmission & distribution (T&D) costs and balancing costs would increase from 2030 onwards: T&D grid costs could rise by €160 billion EUR'17, of which almost half (€70 billion EUR'17) are due to offshore grid cost, while balancing costs could rise by 13 billion EUR'17.Furthermore, this would escalate land-use requirements[21] to approximately 15790 km$^2$ to meet the needs of electricity generation over 2020-2050, which is equivalent to half the area of Belgium. Finally, the additional thermal capacity used to compensate the early NPP closure would heighten air pollution, of which 95% would occur by 2030 (see Section 4.1). Missing the long-term benefits of achieving lower air-pollution levels by 2030 should not be underestimated, as this would be critical point for the EU to display its global leadership role in tackling climate change and limiting global warming to 1.5°C (Intergovernmental Panel on Climate Change, 2018).

---

[21] Land-use costs are difficult to be estimated and are approximated using the geographic footprint.

# 5 Discussion on Full System Costs of Electricity Generation

The most common element to compare technology costs is the levelised cost of electricity (LCOE), which usually incorporates what can be categorised as internal costs over the operational lifetime of the plant, including investment costs, operation and maintenance costs and fuel-cycle costs, taking into account discount rates and inflation (Bustreo et al., 2017; D'haeseleer, 2013; NEA, 2018). Here a distinction is made for nuclear compared to other electricity generation technologies, as decommissioning, waste management and disposal costs are already included in the nuclear LCOE, which are borne by the operator and hence reduce profit. In addition, comparisons based on LCOE, commonly represent *plant-level costs* excluding the influence of *grid-level costs*, i.e. transmission and distribution, balancing and utilisation (or back-up) costs that organise reliable supply (NEA, 2019), which in the end are born either by the grid operator or the end-consumers through taxes or electricity tariffs. Although rough estimates of such costs have been discussed in Section **Error! Reference source not found.Error! Reference source not found.**, in order to properly estimate these costs, they would need to be internalised in the costs of electricity generation technologies and subsequently in the modelling scenarios. This becomes increasingly important as the penetration of variable and intermittent RES in the energy market deepens and as electrification of the economy intensifies; especially when the goal is to decarbonise the power sector at the least cost.

Moreover, except for certain taxes such as the $CO_2$ allowances under the ETS, it has been conventional practice to neglect the external costs, also known as externalities, of power generation technologies (Bickel and Friedrich, 2005; European Commission, 2003). Although accounting for externalities in full system costs is not an uncontroversial topic, as it can be understood as an attempt to reduce human-well-being into a question of euros, electricity generation technologies incur different environmental and social external costs, which are borne by the society at large and should therefore be critical indicators in future policies and decision making. Drawing attention to these understudied issues could facilitate public discussion and policy making by means of integrating the most pressing issues in a meaningful way (NEA, 2018) and in the end positively impact the power sector transformation from a socio-economic perspective.

An example is the share of electricity from nuclear fission, which could increase significantly in the 2050 power mix when incorporating external costs and increasing the decarbonisation ambition (Sangiorgi et al., 2019). Looking beyond 2050, a preliminary analysis suggests that nuclear fission power plants could also act as bridge for fusion deployment by potentially preventing deep and costly changes to the power system, which would thus be ready to accommodate future fusion generation (Cabal et al., 2016). Moreover, internalising externalities could also accelerate fusion penetration, even though a higher decarbonisation ambition would be a more determinant factor in this case (Cabal et al., 2017b; Entler et al., 2018).

Assessments of the external costs of electricity generation technologies have for instance revealed the relatively high external costs of fossil-fuel technologies with respect to human health and climate change impacts, but also revealed the depletion of energy resources and nuclear accidents as the most important parameters driving costs in the case of fission technologies (Alberici et al., 2014; Cabal et al., 2017a). Specifically concerns over risks of accidents and waste management could affect social acceptability and the pace of technology deployment, as is the case in many countries with nuclear energy, even though comparative risk assessments show that health risks are low for nuclear energy (Bruckner et al., 2014;

Hirschberg et al., 2016). It is important to highlight here that, such assessments are highly technology- and country-specific and uncertainties in estimates depend on the availability or lack of historical data, as is the case with some new technologies (e.g. biogas), or future technologies (e.g. some advanced nuclear reactors or fusion technologies). Considering the severity and heterogeneity of energy-related past accidents (Hirschberg et al., 2016), there is a need to improve consistency in analysing health impacts (i.e. related to mortality and morbidity) across all energy-chain stages for all energy technologies, including renewable energy technologies, while taking into consideration not only normal operation but also severe accidents and even hypothetical terrorist threats and health crises. Assessing human health-related impacts, like accidents and toxicity, across all energy technologies and life-cycle stages in a consistent manner would be vital support in managing the political debates around future technological choices and their environmental impacts.

# 6 Conclusions and Policy Implications

By 2050 and compared to 2018, electricity could double its share in the final energy consumption. Electricity will play a crucial role in the decarbonisation and energy transformation in the EU, with the scenarios projecting a fully decarbonised power sector by mid-century. This paper has explored the role of nuclear in the EU policy pathways towards decarbonisation and energy transformation focusing on the power sector and highlighting possible socio-economic implications that could arise when limited social acceptability could virtually exclude the provision of nuclear electricity from the future power mix.

Historical trends reveal that electricity has already become the second most important energy carrier in the EU and the most important one for industry and services. Yet, despite having experienced a large drop in its emissions over the past decades, it still remains EU's largest emitter together with heat. At the same time, the pace of its decarbonisation has been slower than the rate of increase of low-carbon electricity capacity, as a result of the vast majority of new investments being dedicated to new variable and intermittent RES capacities that need four times the installed capacity of nuclear power to produce the same amount of low-carbon electricity. In addition, electricity has held the largest share of energy-related expenditures in EU households. Retail prices grew by almost 20% for households and small and medium-sized industries within the past decade, largely owed to the rise in taxes to support RES and CHP. Electricity production prices were also influenced by RES subsidies that increased their volatility as a result of variations in the generation mix and the marginal costs of generation technologies, which were dominated by the higher coal and natural gas prices.

The EC decarbonisation scenarios reflecting closely the current policy framework, projects electricity to become the second most important source of emissions by 2030 and the share of low-carbon electricity is projected to increase by another 30%. However this would require a remarkable 125% increase in low-carbon power capacity compared to 2017 as a result investments continuing to be dominated by variable and intermittent RES technologies. On the contrary, nuclear capacity would drop by 10% for the first time since 1990, considering the Member States' explicit phase-out policies and ongoing projects at the time of the scenario modelling. The variability of electricity production costs is projected to increase further and retail prices to rise by approximately another 10% compared to 2020. Between 2030 and 2040, a substantial change in electricity infrastructure investments is expected as a large part of the current EU power plants would reach the end of their lifetime operation. The EC 2050 scenarios, project electrification to deepen with increases in decarbonisation ambition and expect a doubling in power capacity investments, continuing to be dominated by variable and intermittent RES technologies, whereas investments in new nuclear power capacities would account for 6%. The overall investments costs are projected to grow even further, particularly in scenarios with a higher deployment of hydrogen and e-fuels that would trigger another rise in electricity prices, even though impacts could be partially mitigated by improved energy efficiency, circular economy and sustainable lifestyle changes to lower consumption.

An alternative scenario where nuclear phase-out policies evolve at a faster pace, would bring the 2050 nuclear power capacity to about 25% of the current levels (36GW), resulting in an additional increase in mid-term investments for variable and intermittent RES and yet-to-be-proven storage technologies, but more importantly for carbon-intensive units (i.e. natural gas and coal). This would thus risk carbon lock-in, stranded assets and potentially heighten EU's energy and material dependence for the coming decades. Compared to a scenario where NPP deployment is not constrained, an early nuclear closure would represent a missed opportunity to cut the power sector $CO_2$ emissions by an extra 17% by 2030 and further boost EU's path towards carbon-free electrification and long-term decarbonisation, putting an upward pressure on EU ETS prices and end-user costs. Indeed, although electricity prices would converge in the long-term, if nuclear baseload phase-out is intensified, power price losses are projected to peak by 2030, which could hamper the competitiveness of electricity against other energy fuels and slow down the electrification of sectors, such as transport. This underlines the urgency for considering long-term economic benefits of lifting restrictions to nuclear power generation that would largely depend on an assumed reduction in the nuclear CAPEX by 2050. Consequently, among the factors that will influence the future role of nuclear in the power mix is not only the ability to be flexible and appropriately located throughout Europe, but also achieve future cost reductions, reactor design and material improvements (e.g. Small Modular Reactors, High Temperature Reactors), as well as design decarbonisation policy actions that recognise the environmental and societal benefits of all low-carbon energy technologies while ensuring a level-playing field across low-carbon technologies. The contribution of nuclear power towards decarbonising the power sector is substantial. Long-term financial support schemes to promote research and technological innovation would allow to leverage its technological benefits and support EU's climate and energy targets. As recommended by authors (Buongiorno et al., 2019), market-oriented economies like the European Union should be able to compete with e.g. in China and Russia to promote technological advancement and investments, with governments adopting an important, yet more limited, role in the development and deployment of nuclear technologies, which could involve establishing sites where companies could test the operation of prototype reactors for regulatory licensing, as well as establish funding programs around prototype testing. The commercial deployment of advanced reactor designs is in the hand of industrial actors and national authorities.

Moreover, decarbonisation exercises evaluating power system costs, assess inconsistently the cost elements across generation technologies and energy-chain stages, and often concentrate on plant-level costs, while disregarding grid-level costs and social and environmental implications, even though these costs are still paid by the society at large. Socio-economic issues that could arise from the energy transition could be with respect to the security of energy supply and import dependency, as well as possible risks of carbon lock-in and issues related to the availability of land, to public acceptance, which could develop as electrification deepens, hindering EU's long-term decarbonisation transition and putting an upwards pressure on end-user costs. A full-system-costs approach could have positive impacts on the long-term socio-economic efficiency of the power sector, while looking beyond 2050, it could facilitate fusion penetration by potentially preventing deep and costly changes to the power system. Incorporating full system costs in future assessments of power-sector transition is especially relevant for technologies most sensitive to policy routes, such as nuclear energy technologies, whose future technological progress and deployment pace is shown to be constrained by social acceptability in many countries, which in turn influence the pace of technology development and cost reduction through economies of scale, thus having the potential to restrict important benefits to be reaped by industries and citizens.

# 7 Acknowledgements

This work was undertaken during the European Commission's Blue Book Traineeship programme, as part of the Euratom Research Unit of the Clean Planet Directorate of the Directorate-General for Research and Innovation. This research did not receive any specific grant from funding agencies in the public, commercial, or not-for-profit sectors.

The views expressed in this work are purely those of the authors and may not under any circumstances be regarded as stating an official position of the European Commission.

The authors would like to thank the following colleagues for providing reports, data and other types of information: a) the Euratom Research Unit of DG RTD, b) the Energy Efficiency and Renewables Unit and the Energy Security, Distribution and Markets Unit of the Joint Research Centre and c) the Economic Analysis and Financial Instruments Unit of the DG Energy.

# 9 Terms and Acronyms

| | |
|---|---|
| BECCS | Bio-energy with carbon capture and storage. |
| CAPEX | Capital Expenditures. |
| CCS | Carbon Capture and Storage technologies. |
| CHP | Combined Heat and Power. |
| EC | European Commission. |
| e-fuels | Are synthetic fuels produced from decarbonised electricity, according to COM(2018) 773. |
| Energy carrier | According to ISO 13600:1997, it is a "substance or phenomenon that can be used to produce mechanical work or heat or to operate chemical or physical processes." It is produced from a primary energy source. |
| Energy resource | Refers to a primary energy resource, a.k.a. natural resource which according to ISO 13600:1997 is a "substance or phenomenon appearing in nature which can be used as input to the technosphere." It has not been not subjected to any human engineered conversion process. |
| ENTSO-E | European Network of Transmission System Operators for Electricity. |
| ESR | Effort Sharing Regulation (ESR) establishes binding annual greenhouse gas emission targets for Member States up to 2030 for most sectors not included in the EU ETS, such as transport, buildings, agriculture and waste. |
| EU ETS | The EU Emissions Trading System (EU ETS) is EU's greenhouse gas emissions trading scheme, covering the sectors of power and heat generation, energy-intensive industry and commercial aviation. |
| EU28 | Also referred to in the text as EU, is the European Union of 28 Member States. |
| Eurostat | European statistics, the statistical office of the European Union. |
| Final energy consumption | The total energy consumed by end users, excluding the consumption of the energy sector itself (in tonnes of oil equivalent). |
| GDP | Gross Domestic Product (US$ or EUR). |
| Gen-IV | Generation IV nuclear reactors. |
| GHG | Greenhouse gases covered by the UNFCCC/Kyoto Protocol, which are carbon dioxide ($CO_2$), methane ($CH_4$), nitrous oxide ($N_2O$), hydrofluorocarbons (HFCs), perfluorocarbons (PFCs), and sulphur hexafluoride ($SF_6$), and nitrogen trifluoride ($NF_3$) (in $CO_2e$). |
| Gross electricity generation | The amount of electricity produced over a period of time, i.e. the electricity measured at the outlet of the main transformers, plus the amount of electricity used in the plant auxiliaries and in the transformers (in Watt hour). |
| Gross inland energy consumption | Abbreviated as gross inland consumption, represents the quantity of energy necessary to satisfy the total energy demand of a country or region (i.e. inland consumption), including the consumption of the energy sector itself, losses during transformation and distribution of energy, and the final consumption by end users. It excludes energy fuels provided to international maritime bunkers (in tonnes of oil equivalent). |
| IEA | International Energy Agency. |
| Import dependency | Also known as energy import rate, is defined as the net energy imports (i.e. imports minus exports) divided by the gross inland energy consumption plus fuel supplied to international maritime bunkers (in %). |
| Installed power capacity | Also used as generation capacity, is the maximum output of electricity that a generator can produce under ideal conditions including the consumption of power stations' auxiliary services and transformers (in Watt). |
| IPCC | Intergovernmental Panel on Climate Change. |
| LCOE | Levelised Cost Of Electricity. |
| Load factor | Electricity generated over maximum potential generation based on net power capacity (in %). |
| LULUCF | Land use, land-use change, and forestry. |
| NEA | Nuclear Energy Agency. |
| Net energy imports | The amount of energy imports minus the amount of energy exports expressed in tonnes of oil equivalent (in tonnes of oil equivalent). |
| Net-generation capacity | Also used as net-power capacity, is the installed power capacity excluding the consumption of power stations' auxiliary services and transformers (in Watt). Gross-electricity generation is defined as the amount of electricity produced over a period of time, i.e. the electricity measured at the outlet of the main transformers, plus the amount of electricity used in the plant auxiliaries and in the transformers (in Watt-hour). |
| NOx | Nitrogen oxides. |
| NPP | Nuclear power plant. |
| OPEX | Operational expenditures. |
| PM | Particulate matter (in thousand tonnes, kt). |
| Primary energy use | Also, primary energy consumption, measures the total energy demand of a country or region, including the consumption of the energy sector itself, losses during transformation and distribution of energy, and the final consumption by end users. It excludes energy fuels used for non-energy purposes (e.g. petroleum used for producing plastics) (in tonnes of oil equivalent). |
| RES | Renewable Energy Sources. |
| SMR | Small modular reactors, a type of nuclear fission reactor that is smaller than conventional reactors, allowing to bypass financial and safety barriers. |
| SOx | Sulphur oxides (in Megatonnes, Mt). |